\documentclass[%
 reprint,
superscriptaddress,
nobibnotes,
 amsmath,amssymb,
 aps,
prb,
citeautoscript,
floatfix,
showkeys
]{revtex4-2}

\usepackage[T1]{fontenc} 
\usepackage[]{graphicx}
\usepackage[utf8]{inputenc}
\usepackage{blindtext}
\usepackage{lmodern}%
\usepackage{hyperref}%
\usepackage{xcolor}%
\usepackage{placeins}
\usepackage{bm}
\usepackage{tabularx}
\usepackage{booktabs}
\usepackage{notes2bib}
\usepackage[normalem]{ulem}
\usepackage{SIunits}
\usepackage[normalem]{ulem}
\usepackage{placeins}

\setcitestyle{super}

\newcommand{\beq}{\begin{equation}\begin{aligned}}
\newcommand{\eeq}{\end{aligned}\end{equation}}

\newcommand{\VIThree}{VI\textsubscript{3}}
\newcommand{\CrIThree}{CrI\textsubscript{3}}

\newcommand{\CGT}{Cr$_2$Ge$_2$Te$_6$}
\newcommand{\CPS}{CrPS$_4$}
\newcommand{\Tc}{\ensuremath{T_{\rm C}}}
\newcommand{\Tn}{\ensuremath{T_{\rm N}}}
\newcommand{\muH}{\ensuremath{\mu_0 H}}
\newcommand{\neel}{\emph{Néel}}
\newcommand{\vdW}{\emph{van der Waals}}

\newcommand{\Vsd}{\ensuremath{V_{\rm {SD}}}}
\newcommand{\VG}{\ensuremath{V_{\rm {G}}}}
\newcommand{\dd}{{\rm d}}

\newcommand{\figref}[2]{\ref{#1}\textsf{\bfseries #2}}
\newcommand{\subpan}[1]{\textsf{\bfseries #1.}}

\newcommand{\ie}{\emph{i.e.},}
\newcommand{\eg}{\emph{e.g.},}

\newcommand{\dqmp}{Department of Quantum Matter Physics, University of Geneva, 24 Quai Ernest Ansermet, CH-1211 Geneva, Switzerland}
\newcommand{\gap}{Department of Applied Physics, University of Geneva, 24 Quai Ernest Ansermet, CH-1211 Geneva, Switzerland}
\newcommand{\KW}{Research Center for Functional Materials, National Institute for Materials Science, 1-1 Namiki, Tsukuba 305-0044, Japan}
\newcommand{\TT}{International Center for Materials Nanoarchitectonics, National Institute for Materials Science, 1-1 Namiki, Tsukuba 305-0044, Japan}
\newcommand{\modena}{Dipartimento di Scienze Fisiche, Informatiche e Matematiche, University of Modena and Reggio Emilia, IT-41125 Modena, Italy}
\newcommand{\cnr}{Centro S3, CNR Istituto Nanoscienze, IT-41125 Modena, Italy}

\definecolor{linkcol}{rgb}{0,0,0.4}
\definecolor{citecol}{rgb}{0.5,0,0}
\definecolor{harvardcrimson}{rgb}{0.79, 0.0, 0.09}
\definecolor{lava}{rgb}{0.81, 0.06, 0.13}

\hypersetup{
	bookmarksopen=true,
	pdftitle="CPS4",
	pdfauthor="Wu et al",
	pdfsubject="", 
	pdfstartview={FitH},    
	pdfmenubar=true, 
	pdfhighlight=/O, 
	colorlinks=true, 
	pdfpagemode=UseNone, 
	pdfpagelayout=SinglePage, 
	pdffitwindow=true, 
	linkcolor=harvardcrimson, 
	citecolor=harvardcrimson, 
	urlcolor=lava
}

	\begin{abstract} 
	Using field-effect transistors (FETs) to explore atomically thin magnetic semiconductors with transport measurements is difficult, because the very narrow bands of most 2D magnetic semiconductors cause carrier localization, preventing transistor operation. Here, we show that exfoliated layers of \CPS\ --a 2D layered antiferromagnetic semiconductor whose bandwidth approaches 1~eV-- allow the realization of FETs that operate properly down to cryogenic temperature. Using these devices, we perform conductance measurements as a function of temperature and magnetic field, to determine the full magnetic phase diagram, which includes a spin-flop and a spin-flip phase. We find that the magnetoconductance depends strongly on gate voltage, reaching values as high as 5000~\% near the threshold for electron conduction. The gate voltage also allows the magnetic states to be tuned, despite the relatively large thickness of the \CPS\ multilayers employed in our study. Our results show the need to employ 2D magnetic semiconductors with sufficiently large bandwidth to realize properly functioning transistors, and identify a candidate material to realize a fully gate-tunable half-metallic conductor.
 
	\end{abstract}

\begin{document}
	
	
	\title{Gate-controlled Magnetotransport and Electrostatic Modulation of\texorpdfstring{\\}{} Magnetism in 2D magnetic semiconductor \texorpdfstring{\CPS}{CrPS4}}%

	\author{Fan Wu}             
    \email{fan.wu@unige.ch}
    \affiliation{\dqmp}
    \affiliation{\gap}
	\author{Marco Gibertini}    
    \affiliation{\modena}
    \affiliation{\cnr}
    \author{Kenji~Watanabe}     
	\affiliation{\KW}
	\author{Takashi~Taniguchi}  
	\affiliation{\TT}
    \author{Ignacio Gutiérrez-Lezama}      
	\author{Nicolas~Ubrig}
	\email{nicolas.ubrig@unige.ch}
    \author{Alberto F. Morpurgo}
    \email{alberto.morpurgo@unige.ch}
    \affiliation{\dqmp}
    \affiliation{\gap}

	\date{\today}

	\maketitle

Atomically thin \vdW\ (vdW) semiconducting magnets (\ie\ 2D magnets) exhibit phenomena resulting from the interplay between semiconducting and magnetic functionalities, absent in other material platforms~\cite{burch_magnetism_2018,gibertini_magnetic_2019,mak_probing_2019}. Recently reported examples include the observation of a strong electric field modulation of the spin-flip field in \CrIThree\ bilayers~\cite{huang_electrical_2018,jiang_controlling_2018}, of the dependence of the luminescence wavelength on the magnetic state of CrSBr~\cite{wilson_interlayer_2021}, and of giant tunneling magnetoresistance of barriers made of 2D magnetic multilayers~\cite{klein_probing_2018,song_giant_2018,wang_very_2018,ghazaryan_magnon-assisted_2018,kim_one_2018,wang_determining_2019}. Theory predicts more phenomena that remain to be observed, such as the possibility to create gate-tunable half-metallic conductors by populating fully spin-polarized bands of 2D semiconductors~\cite{li_half-metallicity_2014,gong_electrically_2018,deng_two-dimensional_2021}. Experiments, however, have been drastically limited by the extremely narrow bandwidths of most 2D magnetic semiconductors explored so far (frequently $\approx~100$~meV)~\cite{wang_electronic_2011,zheng_ab_2019}, which causes these materials to behave very differently from conventional high-quality semiconductors such as Si or GaAs. Indeed, in narrow bandwidth semiconductors Coulomb interactions and disorder potentials easily exceed the kinetic energy of charge carriers, often resulting in their localization. To progress, it is essential to identify 2D semiconducting compounds with the largest possible bandwidth.\\
These considerations are particularly pertinent for the realization of field-effect transistors (FETs) enabling gate-tunable transport measurements well below the magnetic transition temperature \Tc. For the majority of 2D magnets investigated so far (\eg\ Chromium Trihalides~\cite{klein_probing_2018,song_giant_2018,wang_very_2018,ghazaryan_magnon-assisted_2018,kim_one_2018,wang_determining_2019}, MnPS$_3$~\cite{long_persistence_2020}), either no transistor action was observed, or the devices were found to exhibit extremely poor characteristics and to function exclusively for $T$>\Tc. In materials such as \VIThree\ or \CGT\ the bandwidth is slightly larger (a few hundreds meV~\cite{kong_vi3-new_2019} and 0.5~eV~\cite{siberchicot_band_1996,menichetti_electronic_2019}, respectively) and clear transistor action is seen, but even then experiments are drastically limited by the device quality~\cite{wang_electric-field_2018,verzhbitskiy_controlling_2020,zhuo_manipulating_2021,soler-delgado_probing_2022}. Very recently, CrSBr --whose bandwidth exceeds 1.5~eV~\cite{wilson_interlayer_2021,wu_quasi-1d_2022}-- and NiI$_2$ --whose bandwidth is estimated to be 0.8-1~eV~\cite{liu_vapor_2020}-- enabled the first realization of FETs operating down to cryogenic temperature~\cite{wu_quasi-1d_2022,lebedev_electrical_nodate}, confirming that a large bandwidth is indeed key to observe gate-induced in-plane transport at low-temperatures in these systems. However, in neither of these two compounds electrostatic modification of their magnetic phase boundaries has been reported.Currently, therefore, identifying more 2D magnetic semiconductors with a sufficiently large bandwidth to realize FETs enabling the magnetic state of 2D magnetic semiconductors to be probed --and possibly controlled-- remains a key target.\\  
Here, we explore \CPS\ --a 2D semiconducting layered antiferromagnet, whose bandwidth approaches 1~eV-- to show experimentally that FETs based on this material operate properly up to the lowest temperature reached in our experiments ($T$~=~250~mK), and to demonstrate that the magnetic state of the material can be tuned by acting on the gate voltage \VG. Measurements at fixed \VG\ show that the dependence of the device conductance on temperature and applied magnetic field enables the boundaries between the different magnetic states of \CPS\ to be precisely determined. With magnetic field applied perpendicular to the layers, the magnetoconductance exhibits clear features associated to the spin-flop and spin-flip transitions --at respectively $\muH\simeq 0.6$~T and $8$~T, at T~=~2~K-- that shift to smaller fields upon increasing $T$, before eventually disappearing for $T = \Tn$ (the \neel\ temperature of \CPS\ multilayers, $\Tn\simeq35$~K). When the field is applied parallel to the layers, the behavior and magnitude of the magnetoconductance are similar, except for the absence of the spin-flop transition (as expected, since in \CPS\ the easy axis is perpendicular to the layers). The dependence of transport on \VG\ shows the possibility to drastically increase the magnetoconductance, which exceeds 5000~\% as \VG\ approaches the threshold voltage. A shift of the spin-flip and spin-flop fields, as well as of \Tn, is also unambiguously detected, signaling that the magnetic state can be gate-tuned. Besides confirming that large bandwidth 2D magnetic semiconductors enable the magnetic state of 2D magnetic semiconductors to be probed by means of magnetotransport and to be tuned electrostatically, our results identify a first viable experimental candidate for the realization of a gate tunable half-metallic conductor.\\

    \begin{figure}
 		\centering
 		\includegraphics[width=.99\linewidth]{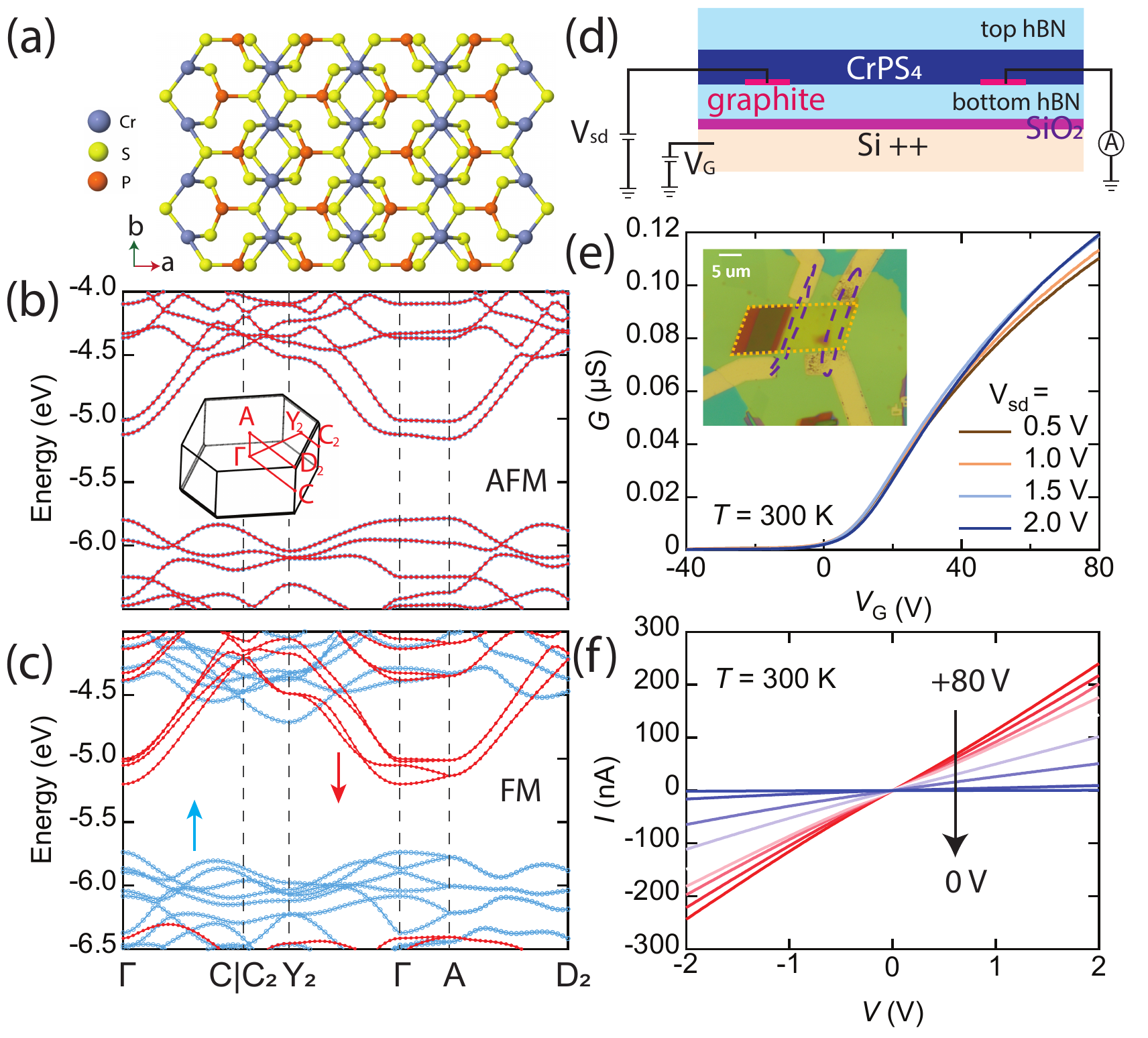}
 		\caption{Properties and device characterization of \CPS\ FETs. \subpan{a} Crystal structure of layered semiconductor \CPS. The blue, yellow and orange balls represent Cr, S and P atoms, respectively. \subpan{b,c} First-principles calculations of the bandstructure of bulk \CPS\ in antiferromagnetic and ferromagnetic ground state, respectively. The same (doubled) unit cell is adopted for both FM and AFM states for ease of comparison. Lines connecting filled red circles and empty blue circles represent spin-up and spin-down states. The inset shows the Brillouin zone and the high-symmetry path along which the bands are computed. \subpan{d} Schematic representation and electrical circuitry of our \CPS\ FETs. \CPS\ is encapsulated with top and bottom hBN and electrical contact to the crystal is provided by inserting few-layer graphene stripes in the stack. \subpan{e} Room temperature transfer curves of a \CPS\ FET under biasing conditions indicated in the legend. The sublinear shape of the curve indicates the presence of a contact resistance (the carrier mobility and subthreshold swing extracted from the transfer curve are 0.2~cm$^2$~V$^{-1}$~s$^{-1}$ and 4 V/decade, respectively). The inset shows a optical microscopic image of the device (the \CPS\ layer is 10 nm thick). The yellow dashed lines delimits the \CPS\ crystal and the graphite electrodes are delimited by the purple dashed lines. \subpan{f} Room temperature IV-characteristics for gate voltages ranging from +80 to 0~V in 10~V steps.}
 		\label{fig:1}
 	\end{figure}
  
	\begin{figure*}[ht]
 		\centering
 		\includegraphics[width=0.8\linewidth]{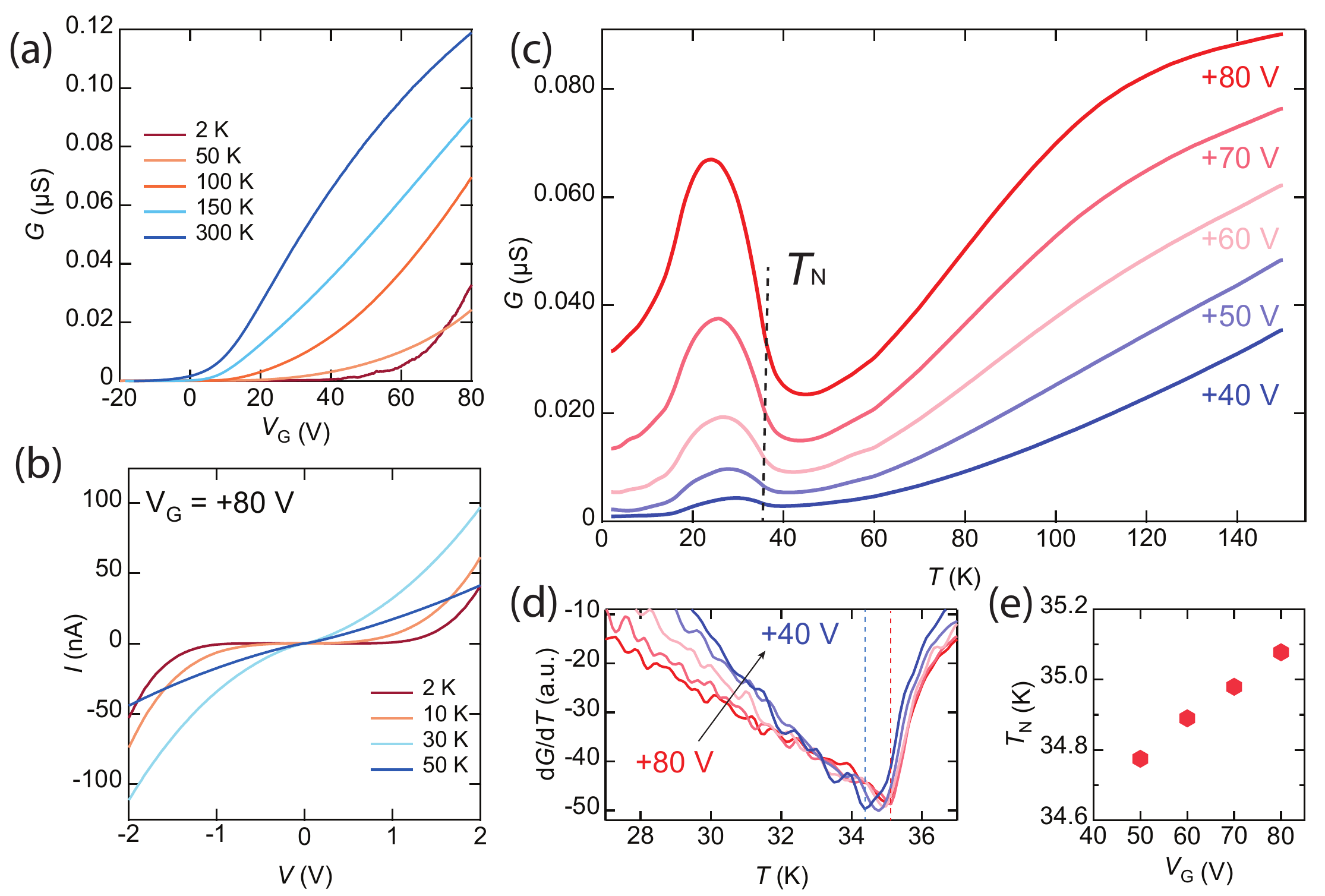}
 		\caption{Temperature-dependent transport in \CPS\ FETs. \subpan{a} Transfer curves of a \CPS\ FET at selected temperatures indicated in the legend. \subpan{b} Low-temperature output characteristics measured at fixed \VG~=~+80~V and selected temperatures of $T$~=~2, 10, 30 and 50~K. \subpan{c} Conductance as a function of temperature measured K at selected \VG\ values (from +40 to +80~V in 10~V steps) with fixed source-drain bias \Vsd~=~2~V to eliminate the effect of the contact resistance present at low bias. The onset of magnetic order coincides with a sharp conductance increase. The dashed line marks the \neel\ temperature $T_N$. \subpan{d} Derivative of the conductance as a function of temperature for the same gate voltages shown in panel \textsf{\bfseries c}. The \neel\ transition temperatures is determined from the position of the minimum in $dG/dT$. Measurements at different $V_G$ between +40 and +80~V exhibit a small but unambiguous shift of transition temperature. \subpan{e} Dependence of the \neel\ temperature \Tn\ on \VG, extracted from panel \textsf{\bfseries d}. For the \VG~=~+40~V trace the transistor is in the subthreshold regime, and the small current prevents a reliable quatitative determination of \Tn.}
 		\label{fig:2}
 	\end{figure*}

\section{Results and Discussion}
 
\CPS\ crystallizes into a monoclinic structure formed by the vdW stacking of distorted CrS$_6$ octahedra and PS$_4$ tetrahedra as shown in Figure~\figref{fig:1}{a}~\cite{louisy_physical_1978,pei_spin_2016,zhuang_density_2016,peng_magnetic_2020,gu_photoluminescent_2020,deng_two-dimensional_2021,calder_magnetic_2020}. It is an air-stable, weakly anisotropic layered antiferromagnet with an easy axis perpendicular to the layers~\cite{peng_magnetic_2020,budko_magnetic_2021,son_air-stable_2021}. The \neel\ temperature --\Tn~=~38~K in bulk crystals-- is somewhat reduced in atomically thin exfoliated multilayers~\cite{peng_magnetic_2020,budko_magnetic_2021,son_air-stable_2021}. Band structure calculations for the antiferromagnetic and ferromagnetic spin configurations of the material (see Figure~\figref{fig:1}{b} and \figref{fig:1}{c}) show that the conduction band is fully spin-polarized in the ferromagnetic state and in individual monolayers, potentially enabling the realization of a fully gate-tunable half-metallic conductors if gate-induced transport at low temperature can be achieved~\cite{deng_two-dimensional_2021}. Most importantly, the calculations also show that the conduction band disperses by approximately 1~eV, which is the main reason for us to select this system to nano-fabricate FET devices.\\
\CPS\ crystals purchased from HQ Graphene are characterized by energy dispersive X-Ray and Raman spectroscopy (see supplementary information), then exfoliated with an adhesive tape, and transferred onto doped silicon substrates covered with a 285~nm thick thermally grown oxide layer (see Figure~\figref{fig:1}{d}). Using by now conventional pick-up and transfer techniques~\cite{wang_one-dimensional_2013}, selected exfoliated layers are placed onto bottom multilayer graphene strips acting as source and drain electrodes, and encapsulated between 20-30~nm thick h-BN crystals. Even though the material is air-stable, exfoliation, transfer and encapsulation are performed in a glove-box with sub-ppm concentration of oxygen and water to maximize the quality of the final device. The structures are then taken out of the glove-box and metal contacts are attached to the graphene strips by means of electron beam lithography, reactive ion etching of the top hBN layer with a CF$_4$/O$_2$ mixture, electron beam evaporation of a Cr-Au film (20/30~nm), and lift-off. An optical microscope image of one of our devices is shown in the inset of Figure~\figref{fig:1}{e}. We realized devices containing multiple transistors on 4 different exfoliated \CPS\ multilayers, with thickness in the 6-11~nm range (as determined by atomic force microscopy), all exhibiting comparable behavior.\\
	\begin{figure*}[ht!]
 		\centering
 		\includegraphics[width=.8\linewidth]{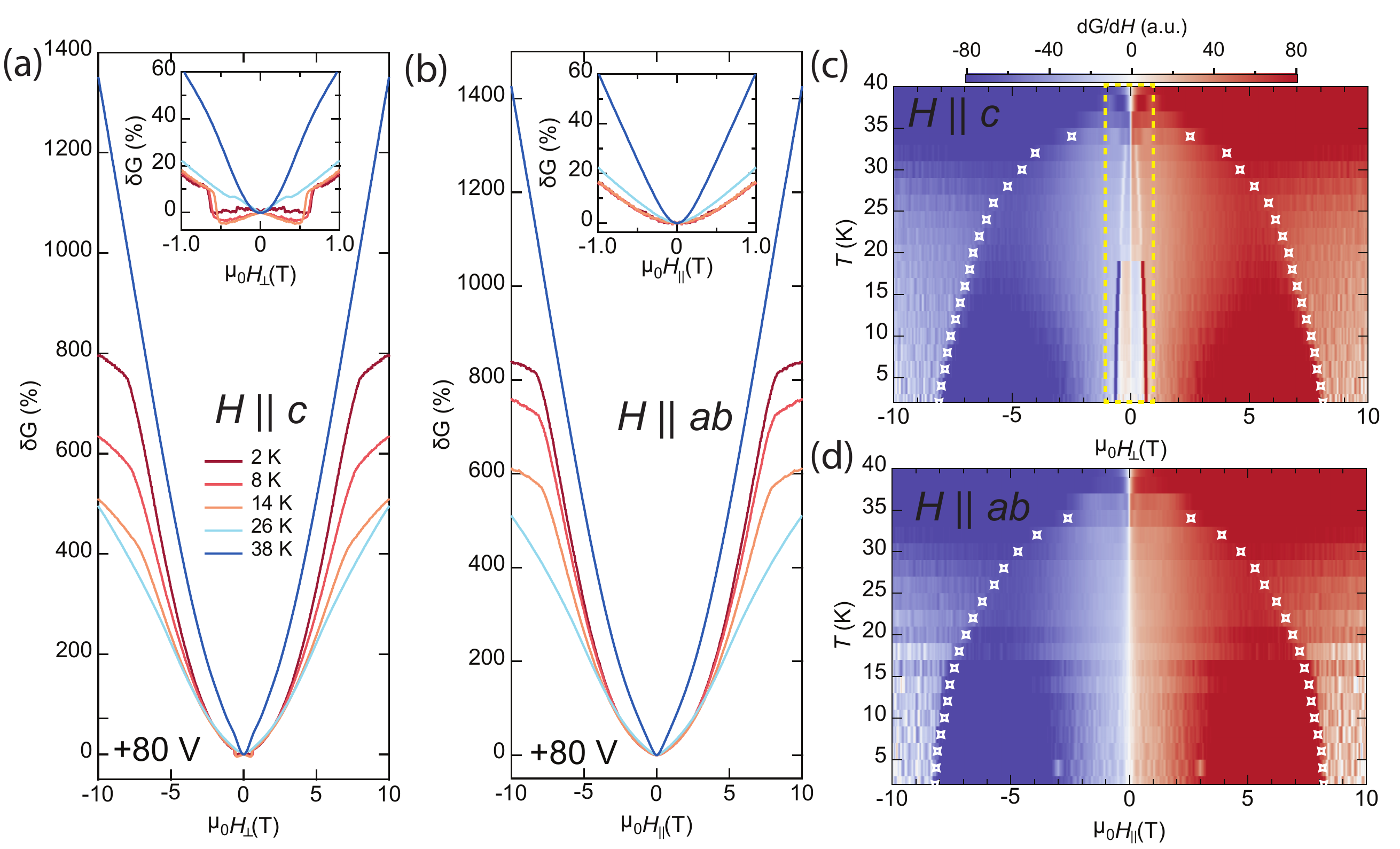}%
 		\caption{Magnetic phase diagram of \CPS. \subpan{a, b} Magnetoconductance ($\delta G = (G(H)-G(0))/G(0)$) measured at $V_G=80$ V, $V_{sd}$ = 2 V and T~=~2, 8, 14, 26 and 38~K, as a function of magnetic field applied respectively perpendicular and parallel to the basal planes. The sharp saturation of the magnetoconductance at high field is due to the spin-flip transition occurring at $\muH \simeq 8$ T at low temperature. The insets zoom-in on the low field range (\muH\ between -1 to 1~T), where the spin-flop transition is visible when the magnetic field is applied perpendicular to the crystalline planes (inset of panel \subpan{a}). \subpan{c, d} Color plot of $\dd G/\dd H$ as a function of temperature and magnetic field applied out-of and in-plane, respectively. The white open symbols mark the spin-flip transition. The yellow dashed box in panel \textsf{\bfseries c} delimits the region around the spin-flop transition.}
 		\label{fig:3}
 	\end{figure*}%
Multilayer \CPS\ FETs exhibit transfer curves (conductance $G$ versus \VG) typical of $n$-type semiconductors, as shown in Figure~\figref{fig:1}{e} with data measured on a 10~nm thick device, whose behavior is representative of that of all transistors investigated in our work (all data shown in the main text have been measured on this same 10 nm device; data from another device are shown in the supplementary information to illustrate that our observations are reproducible). The relatively small threshold ($V_{\rm th}$~=~+10~V) indicates that the Fermi level is located close to the conduction band edge, due to unintentional dopants present in the material. The field-effect (two-probe) mobility estimated from the room-temperature transconductance ($\partial G/\partial\VG$) of this device is 0.2~cm$^2$~V$^{-1}$~s$^{-1}$. Other devices exhibit values reaching up to 1 ~cm$^2$~V$^{-1}$~s$^{-1}$, comparable to values reported in CrSBr and NiI$_2$ transistors, where measurements were performed in a four-probe configuration ~\cite{wu_quasi-1d_2022,lebedev_electrical_nodate}. The output characteristics (source-drain current $I$ versus applied bias $V$, measured for different \VG) are rather linear at room temperature (see Figure~\figref{fig:1}{f}), but upon cooling the behavior changes: the threshold voltages shift to larger positive values (as expected for a doped semiconductor; Figure~\figref{fig:2}{a}), and a pronounced non-linearity of the output characteristics becomes apparent (see Figure~\figref{fig:2}{b}; data measured at \VG~=~+80~V), indicating the presence of a contact resistance. For applied biases \Vsd~>~1.0~V, however, the contacts appear not to influence significantly the measurements, as the extracted electron mobility becomes almost bias independent (see Fig.~S3 in the supplementary information; at low temperature the mobility measured at \Vsd~>~1.0~V ranges between 1-4~cm$^2$V$^{-1}$s$^{-1}$ depending on the device, again comparable to values extracted from four-terminal measurements in CrSBr and NiI$_2$~\cite{wu_quasi-1d_2022,lebedev_electrical_nodate}). We conclude that the quality of our \CPS\ transistors is comparable to that of similar devices recently reported on CrSBr and NiI$_2$, and that it is amply sufficient to allow transport to be measured to temperatures much lower than \Tn, and to be gate-tuned continuously. This conclusion confirms the need to select 2D magnetic semiconductors with sufficiently large bandwidth for the realization of transistors properly functioning at low temperature.\\
    \begin{figure*}[t!]%
 		\centering
 		\includegraphics[width=.8\linewidth]{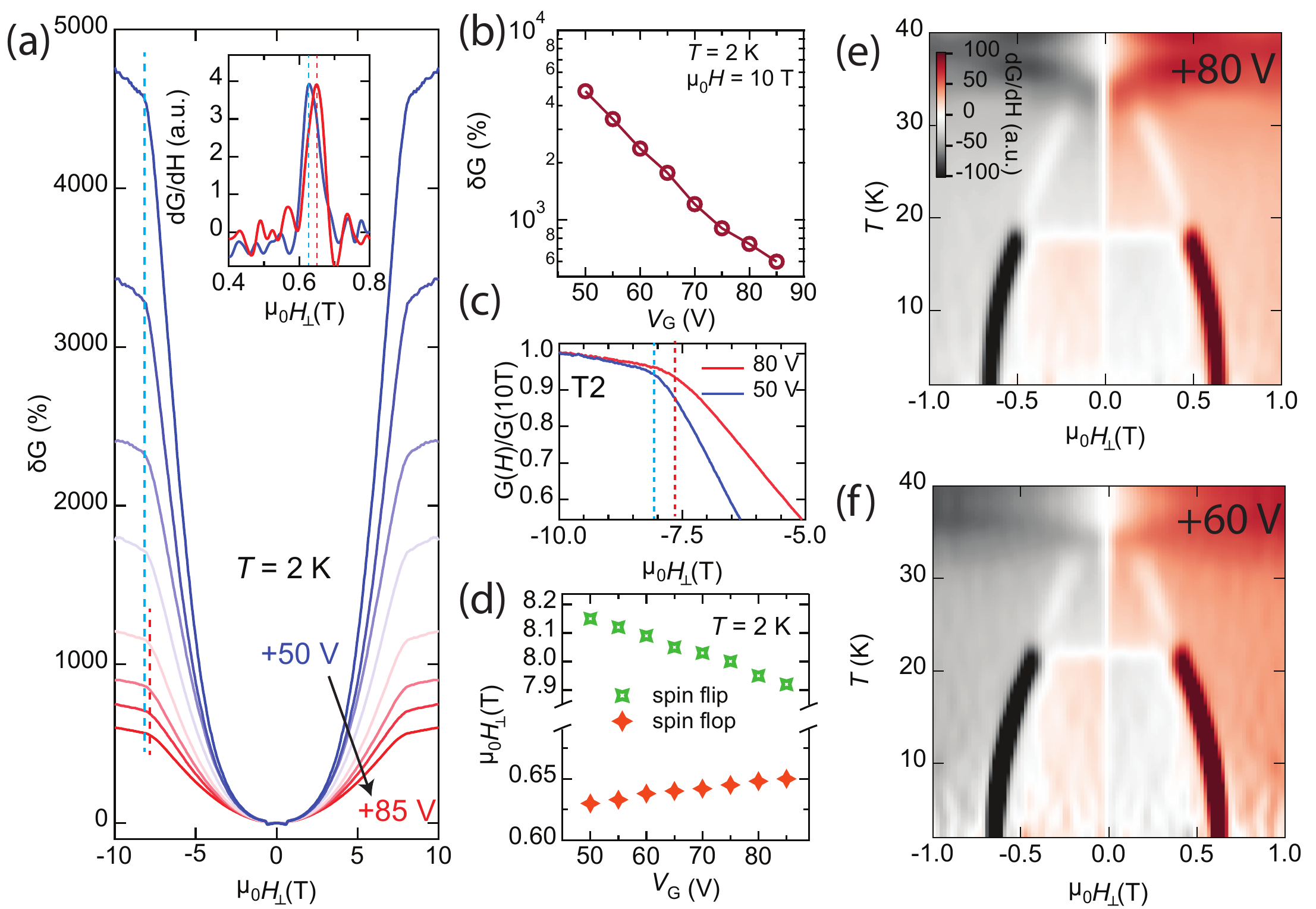}%
 		\caption{Gate modulation of magnetotransport and of the magnetic state of \CPS. \subpan{a} Magnetoconductance measured at \Vsd~=~2~V and $T$~=~2~K for different values of \VG\ between +50~V and +85~V. The blue and red dashed lines indicate the onset of saturation of $\delta G$ occurs at slightly different magnetic field at \VG~=~+50~V and +85~V, showing that the spin-flip transition can be tune electrostatically. The inset shows the derivative of the magnetoconductance in the 0.4 to 0.8~T range, with the peak corresponding to the spin-flop transitions field. Data taken at \VG~=~+50~V (blue line) and +85~V (red line) show that the spin flop transition shifts by approximately 20~mT upon varying \VG. \subpan{b} Gate voltage dependence of the magnetoconductance value at 10~T, showing a nearly exponential dependence of $\delta G$ on \VG. \subpan{c} $G(H)/G(10~T)$ measured at $T$~=~2~K on another device based on a 10~nm \CPS\ exfoliated multilayer, zooming in on the shift of the spin-flip field when increasing \VG\ from 50~V (blue line) to 80~V (red line). \subpan{d} Gate dependence of the spin-flip (open green symbols) and spin-flop (open orange symbols) transition fields. Data measured at $T$~=~2~K. \subpan{e},\subpan{f} Zoom-in on the spin-flop transition region in the color plot of $\dd G/\dd H$ measured at $T$~=~2~K, for \VG~=~+60~V and +80~V, respectively. Besides the shift of the spin-flop field (see inset of panel \subpan{a}), also the details of the magnetoconductance along the phase boundary exhibit a pronounced influence on the gate voltage, with the temperature at which $\dd G/\dd H$ changes sign that varies by nearly 5 K as \VG\ varies from +60~V to +80~V. The effect, whose origin remains to be understood, further illustrates the electrostatic tunability of the magnetic properties of \CPS\ multilayers.}
 		\label{fig:4}
 	\end{figure*}%
The onset of the magnetically ordered state in \CPS\ is clearly visible in the transistor conductance measured as a function of $T$, at fixed \VG. Upon cooling down the devices from room temperature, the conductance decreases slowly (as carriers freeze out into the dopants where they originate from~\cite{Sze_physics_2006}), before exhibiting an abrupt increase as $T$ is lowered just below the \neel\ transition (see Figure~\figref{fig:2}{c}). The abrupt increase results from the transition into the layered antiferromagnetic state that occurs at \Tn~$\simeq$~35~K (as determined by the position of the minimum of $\dd G/\dd T$ in Figure~\figref{fig:2}{c}), slightly lower than the value of T$_N$ in bulk \CPS\ crystals, confirming the trend demonstrated in recent magneto-optical studies that the critical temperature of \CPS\ decreases with decreasing thickness of the crystal~\cite{son_air-stable_2021}. The observed conductance increase below T$_N$ is due to the onset of long-range spin ordering that suppresses disorder associated to spin fluctuations within each \CPS\ layer. At \VG~=~+80~V the effect is sizable, as the conductance increases by more than a factor of 3 before eventually decreasing again and saturating at low temperature.\\
Whether the sizable increase in conductance upon passing from the paramagnetic to the antiferromagnetic state implies that the devices should also exhibit sizable magnetoconductance $\delta G = (G(H)-G(0))/G(0)$ (since an applied magnetic field can modify the magnetic state of the material) is not clear a priori. Large magnetoconductance in magnetic materials commonly occurs when the current flows across regions  with spatially inhomogeneous magnetization (such as domain walls~\cite{gregg_giant_1996,viret_spin_1996,levy_resistivity_1997}, spin valves~\cite{julliere_tunneling_1975,zutic_spintronics_2004}, tunnel barriers made of layered antiferromagnetic insulators~\cite{klein_probing_2018,song_giant_2018,wang_very_2018,ghazaryan_magnon-assisted_2018,kim_one_2018,wang_determining_2019}, etc.). This is not the case in our FETs, since the current in the transistor channel propagates parallel to the \CPS\ layers, and the magnetization of each layer is uniform at low temperatures. Another known mechanism that --by analogy-- may seem relevant for our devices is the one responsible for the magnetoconductance of metallic magnetic multilayers~\cite{baibich_giant_1988,binasch_enhanced_1989,parkin_giant_1991}. When current flows parallel to the layers, the conductance is typically higher in the ferromagnetic state (\ie\ when all layers are magnetized in the same direction) than in the antiferromagnetic one (when adjacent layers are magnetized in opposite directions), because the different scattering rates of majority and minority spin electrons result in a different electron mobility in the two magnetic states~\cite{camley_theory_1989}. This mechanism normally leads to magnetoconductance values well below 100~\% (\ie\ changes that are less than a factor of two), \ie\ smaller than the change that associated to the transition from the paramagnetic to the antiferromagnetic state upon lowering $T$(with \VG~=~+80~V see Figure~\figref{fig:2}{c}). There is therefore no established mechanism suggesting that a very large magnetoconductance should be observed in \CPS\ FETs for $T$~<~\Tn, and indeed, recent measurements on FETs based on CrSBr and NiI$_2$~\cite{wu_quasi-1d_2022,lebedev_electrical_nodate} --which are also 2D semiconducting layered antiferromagnets-- have shown only a modest magnetoconductance (well below a factor of two).\\
Figures~\ref{fig:3} and \ref{fig:4} show that --in contrast to CrSBr and NiI$_2$ transistors-- \CPS\ FETs do exhibit a very large magnetoconductance, which appears upon cooling, as the \neel\ temperature is approached. The large magnetoconductance (see Figures~\figref{fig:3}{a} and \figref{fig:3}{b}) observed above and just below the critical temperature originates from spin fluctuations, because in this temperature range (i.e., in the paramagnetic state or in the antiferromagnetic one, when magnetic order is not yet fully developed) the magnetic susceptibility is largest, so that the applied magnetic field is most effective in polarizing the spins and thereby in reducing spin-induced disorder~\cite{blawat_unusual_2022}. What is however more surprising and unexpected --because far below \Tn\ the magnetic state is robust and spins not easily polarizable-- is that upon lowering the temperature further, the magnetoconductance remains large. Irrespective of whether the magnetic field is applied perpendicular or parallel to the layers, the magnetoconductance reaches values as large as a factor of 50 (\ie\ 5000~\%, seen Figure~~\figref{fig:3}{a} and \figref{fig:3}{b}), for an applied magnetic field \muH~=~10~T, and is gate tunable (the extremely strong, nearly exponential, dependence of $\delta G$ on \VG\ is shown in Figure~\figref{fig:4}{b}). Finding that the large magnetoconductance persists when the field is applied in the plane of \CPS\ --in conjunction with the low electron mobility-- implies that the orbital effects of the magnetic field (as well as the influence of the graphene contacts) are not the reason for the observed large magnetotransport response. Indeed, the experiments indicate that the large magnetoconductance originates from the magnetic-field induced modifications of the magnetic state of \CPS.\\
To illustrate the relationship between magnetoconductance and magnetic state we start by discussing Figure~\figref{fig:3}{c} and \figref{fig:3}{d}, where we plot the derivative of the conductance $\dd G/\dd H$ as a function of field and temperature. For perpendicular fields (Figure~\figref{fig:3}{a} and Figure~\figref{fig:3}{c}), two distinct features are observed. First, for $T$ well below T$_N$ the conductance exhibits a kink and stops increasing rapidly above approximately 8~T (see Figure~\figref{fig:3}{a}). The phenomenon originates from the spin-flip transition, above which all spins fully align along the applied field (in the color plot of $\dd G/\dd H$ in Figure~\figref{fig:3}{c}, the transition --marked by the white diamonds-- manifests itself in a more pronounced way, and can be followed all the way up T$_N$). The second feature is a smaller, but clearly distinct change in magnetoconductance at much lower field ($\muH \simeq \pm$0.6~T at $T$~=~2~K), shown in the inset of Figure~\figref{fig:3}{a}, and clearly apparent in the region delimited by the yellow dashed line in Figure~\figref{fig:3}{c}. It is a manifestation of the spin-flop transition that occurs because \CPS\ is a weakly anisotropic layered antiferromagnet. The color plot of $\dd G/\dd H$ in Figure~\figref{fig:3}{c} illustrate clearly how the spin-flip and spin-flop fields decrease as the temperature is increased, and eventually vanish at approximately 35~K, i.e. the \neel\ temperature $\Tn$ of the material. Similar considerations can be made by looking at Figure~\figref{fig:3}{b} and Figure~\figref{fig:3}{d}, when the magnetic field is applied parallel to the layers, but in that case no spin-flop transition is observed (as expected, since the easy axis of CrPS$_4$ is perpendicular to the layers). Measuring the magnetoconductance of our FETs, therefore, allows us to trace the complete magnetic phase diagram of \CPS\ multilayers, and even to identify features that are often complex to detect experimentally in atomically thin layers (such as the spin-flop transition, which has not been detected in any exfoliated 2D magnet by means of magneto-optics experiments, the technique possibly the most commonly employed to study magnetism in these systems).\\
Having established the ability to map the magnetic phase diagram through magnetoconductance measurements, we analyze whether the different magnetic states can be influenced electrostatically, by varying the voltage applied to the gate. A strong effect is only expected in devices realized with crystals that are one or few layers thick because electrostatic screening forces the charge accumulated in the transistor channel to reside exclusively in the first few layers next to the gate electrode (in the range of accumulated electron densities --a few 10$^{12}$ cm$^{-2}$-- the screening length is approximately 1 nm, which is why most charges are accumulated in the first CrPS$_4$ layer next to the gate). Nevertheless --despite our devices being 10-to-20 monolayer thick-- magnetoconductance measurements unambiguously show that systematic gate-induced changes in the magnetic state are present. To demonstrate this conclusion, we focus on specific points of the boundaries between the different magnetic phases that can be precisely determined, and look at their evolution upon varying \VG.\\
First clear evidence for the modification of the \CPS\ phase boundaries by the electrostatic gate is seen in the critical temperature \Tn\ at zero applied magnetic field, which experimentally can be precisely determined by looking at the position of the minimum in $\dd G/\dd T$  (Figure~\figref{fig:2}{d}). Upon varying \VG\ from +50~V to +80~V, the minimum in $\dd G/\dd T$ --and hence \Tn-- increases by approximately 0.5~K (Figure~\figref{fig:2}{e}; in the different devices that we measured the change ranges between 0.5 to 1~K, see for instance Fig.~S5 in the supplementary information for data from another transistor. Varying the gate voltage also affects the spin-flip transition field, which decreases by about 0.3~T when the gate is changed in approximately the same range at $T$~=~2~K (see the blue and red dashed line in Figure~\figref{fig:4}{a} and {c}), and modulates the spin-flop transition as well (compare Figure~\figref{fig:4}{e} and \figref{fig:4}{f}): the magnetoconductance along the phase boundary changes and the spin-flop field increases by approximately 20~mT as \VG\ is increased from +50~V to +85~V (see the inset of Figure~\figref{fig:4}{a}). The evolution of both transitions as a function of gate voltage is summarized in Figure~\figref{fig:4}{d}. Since, at the simplest level, the spin-flip field is proportional to the interlayer exchange interaction $J$, and the spin-flop field to $\sqrt{JK}$ (where $K$ is the magnetic anisotropy)~\cite{rosler_magnetic_2004}, finding that these two quantities change in opposite directions (the first decreases whereas the second increase as \VG\ is increased) implies that both $J$ and $K$ are affected by the gate voltage.\\
Finding that \CPS\ FETs allow controlling the magnetic state electrostatically and probing the resulting changes by transport measurements is worth emphasizing because the only 2D magnetic semiconductors on which continuous FET operation could be achieved at low temperature so far (CrSBr~\cite{wu_quasi-1d_2022} and NiI$_2$~\cite{lebedev_electrical_nodate}) did not exhibit any modulation of the magnetic phase boundaries upon varying the gate voltage. Indeed, among all 2D magnetic semiconductors studied, only in mono and bilayers of \CrIThree\ a continuous gate-tuning of the magnetic state has been observed, by means of magneto-optical techniques~\cite{jiang_controlling_2018,huang_electrical_2018}. In that case, all layers were affected by the gate-accumulated (reaching up to 10$^{13}$~cm$^{-2}$) --owing to the atomic thickness of the material-- and for monolayer \CrIThree, a 10~K gate-induced change in critical temperature was reported~\cite{jiang_controlling_2018}. In our transistors the effect is somewhat smaller, probably because the charge is accumulated predominantly in the individual layer closest to the gate electrode, but the \CPS\ multilayers are 10-to-20 layers thick, and we expect that in devices based on \CPS\ mono or bilayers a stronger dependence of the magnetic state on gate voltage will be observed. Exploring how the magnetic state depends on thickness and how thin multilayers are influenced by charge accumulation is certainly worth addressing in future work, to reveal the microscopic mechanism through which charge accumulation modifies magnetism that is currently unknown.\\

\section{Conclusion}
The ability to accumulate electrostatically charge carriers at the surface of 2D magnetic semiconductors to induce electronic transport and control the magnetic state discloses opportunities for future experiments targeting unexplored phenomena. Identifying the microscopic mechanism responsible for the interplay between transport electrons and the magnetic state reported here --causing the magnetoconductance to reach values up to 5000~\%-- provides a first clear objective. More in general, the occurrence of gate-induced low-temperature transport in \CPS\ enables a much broader variety of experiments to be done, for instance in multi-terminal devices (\eg\ to search for the anomalous Hall effect and investigate its gate dependence), in devices based on few-layer crystals (where the influence of gating on the magnetic state is expected to be much stronger), and in monolayers (where it will be possible to realize gate-tunable half-metallic conductors). Of great interest is also the implementation of local gates, to control the magnetic state locally in different parts of the device (to explore, for instance, whether the presence of interfaces between different magnetic states within the transistor channel allows the magnetoconductance to be increased even further). It should be realized, however, that the scope of possibilities is much broader, because \CPS\ is only the first large-bandwidth magnetic 2D semiconductor that enables the realization of transistors properly functioning at low temperature. We envision that many more 2D magnetic semiconductors with bandwidth comparable to (or larger than) that of \CPS\ will be introduced in the future, hosting a rich variety of magnetic states, providing experimental access to an even much broader variety of interesting physical phenomena.

    \section*{Experimental section}
\emph{Sample fabrication}. \CPS\ multilayers were obtained by exfoliating bulk crystals, purchased from HQ Graphene, in a Nitrogen gas-filled glove box, with sub-ppm oxygen and water concentration to exclude any degradation and maximize the quality of the final device. The multilayer thickness was determined by means of atomic force microscopy. The multilayers were contacted electrically with  few-layer graphene stripes and encapsulated with exfoliated hBN layers, employing a conventional pick-up and release technique based on PC/PDMS polymer stacks placed on glass slides. A total of 4 devices have been fabricated and the graphene stripes contacted with metallic electrodes (Cr/Au: 10/40 nm), using a conventional process based on electron-beam lithography, reactive ion etching, electron-beam evaporation, and lift-off.\\
\emph{Transport measurements}. All transport measurements were performed in a Heliox 3He insert system (Oxford Instruments) equipped with a 15~T superconducting magnet.  A Keithley 2400 source/measure was used to apply the gate voltage to the transistors. The bias voltage was applied using a homemade low-noise source. The current and voltage signals were amplified with homemade low-noise electronics and the amplified signals were recorded with an Agilent 34410A digital multimeter unit.\\
\emph{DFT calculations}. First-principles simulations were performed using the Quantum ESPRESSO distribution~\cite{giannozzi_quantum_2009,giannozzi_advanced_2017}. To account for van der Waals interactions between the layers, the spin-polarised extension\cite{thonhauser_spin_2015} of the revised vdw-DF-cx exchange-correlation functional~\cite{dion_van_2004,lee_higher-accuracy_2010} was adopted, which gives lattice parameters in good agreement with experiments. The Brillouin zone was sampled with a $6\times6\times3$ $\Gamma$-centered Monkhorst-Pack grid, considering a doubled unit cell in the vertical direction (containing 24 atoms) both in the ferro- and antiferro-magnetic configurations. Pseudopotentials were taken from the Standard Solid-State Pseudopotential (SSSP) accuracy library (v1.0)~\cite{prandini_precision_2018} with cutoffs of 40~Ry and 320~Ry for wave functions and density, respectively. Atomic positions and lattice parameters were relaxed, reducing forces and stresses below thresholds of 3~meV/\AA\ and 0.5~kbar, respectively. The band structure was computed along a high-symmetry path obtained by selecting the most interesting directions provided by SeekPath~\cite{hinuma_band_2017}.

\section*{Acknowledgements}
	The authors gratefully acknowledge Alexandre Ferreira for continuous and valuable technical support. AFM gratefully acknowledges the Swiss National  Science Foundation and the EU Graphene Flagship project for support. MG acknowledges support from the Italian Ministry for University and Research through the Levi-Montalcini program. KW and TT acknowledge support from the Elemental Strategy Initiative conducted by the MEXT, Japan (Grant Number JPMXP0112101001) and JSPS KAKENHI (Grant Numbers 19H05790, 20H00354 and 21H05233).

%

\end{document}